\documentclass{aa}

\newcommand{\gapeq}{\mbox{$~\stackrel{\scriptstyle >}{\scriptstyle \sim}~$}}
\newcommand{\lsol}{L$_{\odot}$}
\newcommand{\lfir}{$L_{\rm FIR}$}
\newcommand{\lb}{$L_{\rm B}$}
\newcommand{\msol}{M$_{\odot}$}

\newcommand{\hi}{H\,{\small{\sc I}}}
\newcommand{\hii}{H\,{\small{\sc II}}}

\newcommand{\esn}{$\dot E^{\rm tot}_{\rm SN}$}
\newcommand{\edota}{$\dot E^{\rm tot}_{\rm A}$}

\newcommand{\mk}{$M_{\rm K}$}
\newcommand{\ergscm}{ergs s$^{-1}$ cm$^{-2}$}

\usepackage{graphics}
\usepackage{psfig}

\begin{document}

\hyphenation{con-sti-tu-ents there-by mar-gin-al-ly stat-is-tics}

\title{Dependence of radio halo properties on \\ star formation activity 
and galaxy mass}

\titlerunning{Properties of radio halos}
\authorrunning{M. Dahlem et al.}

\author{Michael Dahlem\inst{1}
\and 
Ute Lisenfeld\inst{2,3}
\and
J\"orn Rossa\inst{4}
}

\institute{CSIRO/ATNF, Paul Wild Observatory, Locked Bag 194,
Narrabri NSW 2390, Australia
\and
Dept. F\' isica Te\' orica y del Cosmos, Facultad de Ciencias, 
Universidad de Granada, 18071 Granada, Spain
\and
Instituto de Astrof\' isica de Andaluc\' ia, CSIC, Apdo. 3004, 18080 
Granada, Spain
\and
Space Telescope Science Institute, 3700 San Martin Drive, Baltimore,
MD 21218, USA}

\offprints{M.\,D; Michael.Dahlem@csiro.au}

\date{Received 30 December 2005 / Accepted 29 June 2006}

\abstract{
We investigate the relation between the existence and size of radio 
halos, which are believed to be created by star formation (SF) related 
energy input into the interstellar medium, and other galaxy properties, 
most importantly star formation activity and galaxy mass.
Based on radio continuum and H$\alpha$ observations of a sample of 
seven late-type spiral galaxies we find a direct, linear correlation 
of the radial extent of gaseous halos on the size of the actively 
star-forming parts of the galaxy disks.
Data of a larger sample of 22 galaxies indicate that the threshold 
energy input rate into the disk ISM per unit surface area for the 
creation of a gaseous halo depends on the mass surface density of 
the galaxy, in the sense that a higher threshold must be surpassed 
for galaxies with a higher surface density.
Because of the good prediction of the existence of a radio halo from 
these two parameters, we conclude that they are important, albeit
not the only contributors.
The compactness of the SF-related energy input is also found to be a 
relevant factor. Galaxies with relatively compact SF distributions 
are more likely to have gaseous halos than others with more widespread 
SF activity.
These results quantify the so-called ``break-out'' condition for
matter to escape from galaxy disks, as used in all current models
of the interstellar medium and first defined by Norman \& Ikeuchi
(1989).
}
\maketitle

\section{Introduction}
\label{par:intro}

\begin{table*}[ht!]
\begin{flushleft}
\leavevmode
\caption{Galaxy Sample}
\label{tab:sample}
\begin{tabular}{lccccc}
\noalign{\hrule\smallskip}
\noalign{\hrule\smallskip}
  & NGC\,1808 &  M\,82  & NGC\,4666 & NGC\,4700 & NGC\,7090 \\
\noalign{\hrule\smallskip}
$\alpha$(2000)  & 05:07:42.3 & 09:55:52.2 
  & 12:45:08.3 & 12:49:07.6 & 21:36:28.6 \\
$\delta$(2000)  & --37:30:47 & +69:40:47 
  & --00:27:51 & --11:24:47 & --54:33:24 \\
$v_{\rm hel}$ (km s$^{-1}$) & 1000 & 203 & 1520 & 1404 & 849 \\
{\it D} (Mpc) & 10.9 & 3.2 & 26.4 & 25.5 & 11.7 \\
{\it PA} ($^\circ$) & 320 & 60 & 225 & 48 & 130 \\
{\it i} ($^\circ$) & 60 & 90 & 78 & 90 & 90 \\
\noalign{\smallskip\hrule}
\end{tabular}
\end{flushleft}
The entries in the rows of Table~\protect\ref{tab:sample}\ are: \\
$\alpha,\delta$(2000): J2000 equinox centre coordinates \\
$v_{\rm hel}$: Heliocentric velocity, from Koribalski et al. (1993;
NGC\,1808), de Vaucouleurs et al. (1991; M\,82), Walter et al. (2004; 
NGC\,4666) and Dahlem et al. (2005; NGC\,4700 and NGC\,7090) \\
{\it D}: Distances are based on $H_\circ$ = 75 km s$^{-1}$ Mpc$^{-1}$
and a virgocentric infall velocity of 300 km s$^{-1}$ (cf. Dahlem et
al. 2001). \\
{\it i:} Inclination angle \\
\end{table*}

The investigation of gaseous halos in late-type spiral galaxies is 
closely linked to studies of the galaxies' chemical evolution. 
Halos are potential mediators of metal redistribution via disk-halo 
interactions (e.g. Bregman 1980).
If escape velocity is reached, for example in galactic winds,
outflowing halo gas influences the evolution of the intergalactic 
medium by means of metal and energy injection, thus adding to
the potential importance of gaseous halos on cosmological scales
(e.g. Heckman 2005).

Over the last decade it has become clear that spiral galaxies
with high-mass star formation (SF) have gaseous halos (e.g., Heckman,
Armus, Miley 1990; Lehnert and Heckman 1995; Dahlem 1997; Rossa
and Dettmar 2003a). 
In fact, we have found evidence (Dahlem et al. 1998; Dahlem et al. 
2001) that almost {\it all} far-infrared (FIR) ``warm'' edge-on 
spirals (i.e., with 60 $\mu$m to 100 $\mu$m FIR flux ratios 
$f_{60}/f_{100} \gapeq 0.4$) have gaseous halos. The only remaining 
exceptions from this rule are galaxies for which no firm statement 
can be made due to the lack of data of sufficient quality.

These gaseous halos comprise all known phases of the interstellar
medium (ISM) previously found in the disks (e.g., Dahlem 1997).
Optical (H$\alpha$) emission was detected in M\,82 already
by Lynds and Sandage (1963). Radio detections came later, starting
with NGC\,4631 (Ekers and Sancisi 1977). Yet later, X-ray emission was
detected from halos (e.g., Watson et al. 1984). 

Further progress in this field of research has been made by studies
in various wavebands, such as optical H$\alpha$ imagery (Rossa
and Dettmar 2000, 2003a,b; see also the work by Dettmar 1992 
and Rand 1996) and X-ray imaging spectroscopy (e.g., Dahlem et al.
1998, Pietsch et al. 2000, Strickland et al. 2004a,b and references 
therein; T\"ullmann et al. 2006).
\hi\ was detected in halos around spirals (e.g., Fraternali et al. 
2004 and references therein; Boomsma et al. 2005) and, most recently, 
diffuse ultra-violet (UV) emission was found from the starburst 
outflows of NGC\,253 and M\,82 (Hoopes et al. 2005).

Radio synchrotron emission from relativistic electrons is the most 
extended and pervasive component of these halos, which makes them 
most easily detectable.
Thus, recent attempts to detect radio emission from extraplanar gas 
in actively star-forming galaxies have been quite successful (e.g.
Irwin et al. 1999; Dahlem et al. 2001). 
Nevertheless, a debate is still continuing about how such halos 
are created and energetically maintained (Dahlem 1997).
In a first attempt to establish general rules of behaviour for
gaseous halos of spirals, we (Dahlem, Lisenfeld, Golla 1995;
hereafter DLG95) investigated the dependence of the properties
of radio halos on the level of SF in the underlying disks of 
NGC\,891 and NGC\,4631. At the time, these were the only two 
galaxies for which we had data of sufficient quality to conduct 
such studies. 1.49 GHz radio continuum images were used to
measure the properties of the radio halos of both galaxies.
The most suitable tool to determine the radius inside which 
high-mass SF is occurring, $r_{\rm SF}$, are FIR 
data. However, only very few images with sufficient angular 
resolution exist (in particular those obtained with the
IRAS CPC detector; see van Driel et al. 1993). In the case of
NGC\,4631 we substituted the FIR data with an H$\alpha$
image.

As part of a small radio continuum survey of FIR-warm edge-on 
spiral galaxies, radio images of several more suitable galaxies 
were obtained (Dahlem et al. 2001, 2005). 
Here we present an analysis of those radio images with
sufficient resolution for detailed studies of halo
properties. We compare these with H$\alpha$ emission line 
images and/or other tracers of SF in galaxy disks, to 
investigate whether these galaxies show a similar dependence 
of their radio halo properties on the level of activity in 
the underlying disks as NGC\,891 and NGC\,4631 (DLG95). 
To ensure that processes related to high-mass SF dominate the 
energy balance of the ISM in these galaxies, systems with known 
luminous active galactic nuclei (AGNs) and closely interacting 
galaxies were excluded (see Dahlem et al. 2001).

\section{Database} 
\label{par:data}

Here we concentrate on radio continuum and H$\alpha$ halos, but
different gaseous phases are normally found associated with each 
other (Dahlem 1997; Veilleux et al. 2005; T\"ullmann et al. 2006). 
Therefore, it is very likely that galaxies with either radio 
continuum or H$\alpha$ halos in general have multi-phase gaseous 
halos, including all the phases of the ISM mentioned above.

\subsection{Sample selection}
\label{par:select}

Only galaxies with known extraplanar emission both in the radio
continuum and optical emission lines were considered. Also, these
galaxies were selected to be neither closely interacting (and 
thus not tidally disturbed) nor to host a dominant AGN and are
therefore suitable for studying the dependence of gaseous halo 
properties on the SF activity in the underlying disks. 
The available datasets must allow the measurement of at least the
following three quantities: 1. The maximum radius at which a spiral
galaxy actively forms stars, $r_{\rm SF}$; 2. the maximum radius at
which nonthermal radio continuum emission from cosmic rays is
observed in their disks, $r_{\rm CR}$; 3. the maximum radius at
which synchtrotron radio continuum emission is detected in the
halo, $r_{\rm halo}$.
These selection criteria severely limit the number of suitable 
targets. 
At present the list of galaxies fulfilling all selection criteria, 
for which we have suitable datasets (in addition to NGC\,891 and 
NGC\,4631; DLG95), comprises five new objects; their basic properties 
are collected in Table~\ref{tab:sample}. For these galaxies the 
following observational datasets are used. 
All data are available to us in electronic, reduced form, except 
the radio continuum data for NGC\,1808 by Saikia et al. (1990). 
For this galaxy the derived radii were taken from the literature.

\subsection{Radio continuum observations}

Low-frequency (1.4 GHz) radio continuum observations are used to 
trace cosmic-ray (CR) electrons in halos around the disks of the 
sample galaxies. 

The radio data analysed here have all been published before. 
The radio data of NGC\,1808 were published by Saikia et al. (1990), 
the data of M\,82 (=NGC\,3034) were obtained by Reuter et al. 
(1992).
The radio image of NGC\,4666 is from Dahlem et al. (1997).
The 1.4 GHz maps of NGC\,4700 and NGC\,7090 are from Dahlem et al. 
(2001). 
The observations and their calibration and data reduction are all
described in the respective papers.

\subsection{H$\alpha$ images}

H$\alpha$ imagery is used as a tracer of active SF in the disks
of galaxies. Almost all H$\alpha$ images used for comparison with 
the radio continuum data have been taken from previous studies.
The H$\alpha$ image of NGC\,1808 was published by Jim\' enez-Bail\' on 
et al. (2005).
M. Lehnert kindly provided us with a wide-field H$\alpha$ image 
of M\,82 (Lehnert et al. 1999).
The H$\alpha$ frame of NGC\,3175 is from Ryder and Dopita (1993).
The NGC\,4666 H$\alpha$ image was originally published by Lehnert 
and Heckman (1995) and has already been used by us earlier (Dahlem 
et al. 1997).
The H$\alpha$ images of NGC\,4700 and NGC\,7090 originate from 
Rossa and Dettmar (2003a,b).

\begin{figure}
\resizebox{1.0\hsize}{!}{\includegraphics{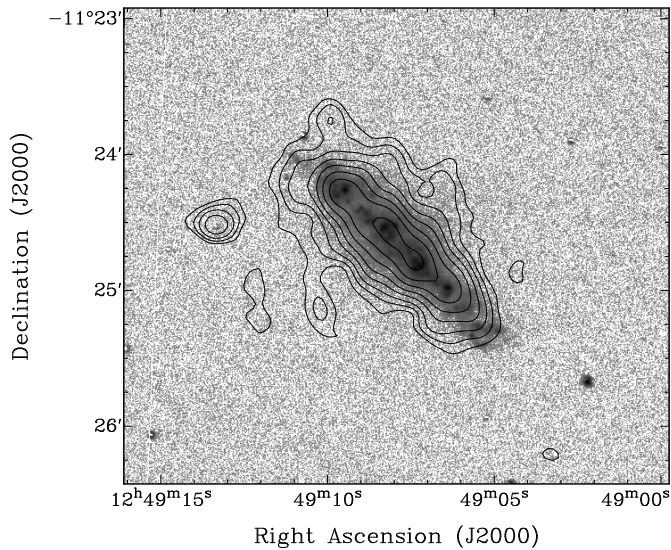}}
\caption{VLA 1.49 GHz radio continuum image overlaid on our H$\alpha$
image of NGC\,4700. Contours range from -0.2, -0.14, 0.14 (2.5-$\sigma$), 
0.2, 0.28, ..., 1.6 mJy beam$^{-1}$, spaced by factors of $\sqrt{2}$, 
while the logarithmic H$\alpha$ grey scales range from 0 to the maximum 
observed surface brightness. The resolution (full width at half maximum) 
of the radio image is $13''$ (1.6 kpc).
}
\label{fig:n4700over}
\end{figure}

\subsection{Additional data}

In addition we include some galaxies in the present study for which 
the measurement of the three radii $r_{\rm halo}$, $r_{\rm SF}$ and 
$r_{\rm CR}$ (see Sec.~\ref{par:select}) is not possible from the
available images. 
Still, observations of these galaxies can be useful and are added 
here (once the existence of a gaseous halo is established) to
investigate the dependence of their halo properties on some 
global galaxy parameters, such as e.g. their average energy input
rates and galaxy mass.

To this end we use additional 1.4 GHz radio continuum images of 
NGC\,1406, NGC\,3175 and NGC\,7462 from Dahlem et al. (2001).
Furthermore data of three galaxies from the sample by Irwin et al. 
(1999) are used. Their analysis of the radio emission is different 
from ours (Dahlem et al. 2001) and not easily comparable. Therefore, 
we only include here galaxies from their sample {\it not} showing 
evidence of emission beyond the modelled thin disks, which is a 
robust criterion similar to the one used by us in the rest of the 
sample.
An H$\alpha$ image of NGC\,7462 from Rossa and Dettmar (2003a,b) 
was also used.

\section{Results}
\label{par:results}

\begin{figure}
\resizebox{1.0\hsize}{!}{\includegraphics{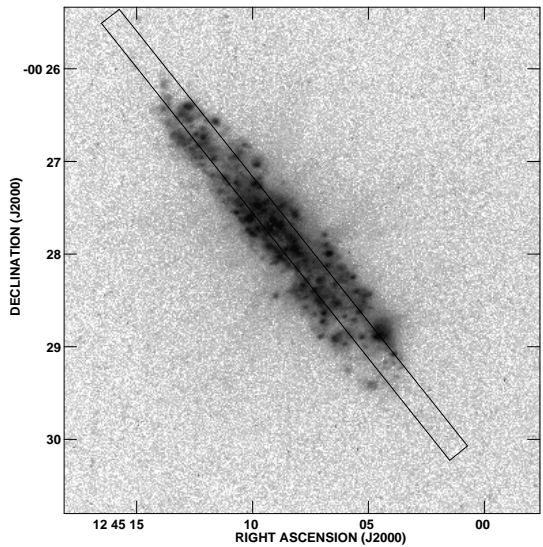}}
\caption{
Placement of a $12''$-wide cut through the H$\alpha$ emission 
distribution of NGC\,4666 along the disk plane, indicating the area 
over which the cut displayed in Fig.~\protect\ref{fig:n4666z0cut}\
was determined. See Table~\protect\ref{tab:cuts}\ for details.
}
\label{fig:n4666haz0}
\end{figure}

\begin{table*}[ht!]
\begin{flushleft}
\leavevmode
\caption{Parameters for Cuts}
\label{tab:cuts}
\begin{tabular}{lcccccccc}
\noalign{\hrule\smallskip}
\noalign{\hrule\smallskip}
~~~(1) &  (2)  &  (3) & (4) & (5) & (6) & \multispan{3}{~~(7)}{ } \\
Object     &   Band  & Location & {\it HPBW} & Width & Measurements 
  & \multispan{3}{~~~~}{$z$-offset}{~~~} \\
           &         & Disk/Halo & (~$''$) & (~$''$) & Averaged 
  & (~$''$) & ({\it HPBW}) & (kpc) \\
\noalign{\hrule\smallskip}
NGC\,1808 &   1.4GHz  & Disk &  0.6 &  5 & 2 &  0 &  0   &  0  \\
          &   1.4GHz  & Halo &  1   &  5 & 2 & 10 &  10  & 0.5 \\
          & H$\alpha$ & Disk &  1   &  5 & 2 &  0 &  0   &  0  \\
M\,82     &   1.4GHz  & Disk &  3.8 &  5 & 2 &  0 &  0   &  0  \\
          &   1.4GHz  & Halo &  3.8 &  5 & 2 & 30 & 7.89 & 0.5 \\
          & H$\alpha$ & Disk &  1   &  5 & 2 &  0 &  0   &  0  \\
NGC\,4666 &   1.4GHz  & Disk & 14   & 12 & 2 &  0 &  0   &  0  \\
          &   1.4GHz  & Halo & 14   & 12 & 4 & 32 & 2.29 & 4.1 \\
          & H$\alpha$ & Disk &  1.9 & 12 & 2 &  0 &  0   &  0  \\
NGC\,4700 &   1.4GHz  & Disk & 13   & 12 & 2 &  0 &  0   &  0  \\
          &   1.4GHz  & Halo & 13   & 12 & 3 & 16 & 1.23 & 2.0 \\
          & H$\alpha$ & Disk &  1   & 12 & 2 &  0 &  0   &  0  \\
NGC\,7090 &   1.4GHz  & Disk & 14   & 30 & 2 &  0 &  0   &  0  \\
          &   1.4GHz  & Halo & 14   & 30 & 3 & 60 & 1.74 & 3.4 \\
          & H$\alpha$ & Disk &  1   & 30 & 2 &  0 &  0   &  0  \\
\noalign{\smallskip\hrule}
\end{tabular}
\end{flushleft}
The entries in the columns of Table~\protect\ref{tab:cuts}\ are: \\
(1) Object name; (2) Observing waveband; (3) Location of cut(s) from 
which measurements were obtained; (4) Angular resolution of data used;
(5) Width of the cut over which datapoints were averaged; (6) Number 
of measurements used to derive mean value; (7) Offset of cuts from 
galaxy disks, in units of arcsec, half power beamwidths and kpc.
\end{table*}

In the publications from which the data were taken most of the
basic emission properties of the sample galaxies are listed 
(e.g., total flux densities, centre positions, etc.).
An example of a (previously unpublished) dataset used here is 
provided in Fig.~\ref{fig:n4700over}, where we display a contour
overlay of a new VLA 1.49 GHz continuum map of NGC\,4700 on our 
H$\alpha$ line image. One can see directly related radio continuum 
and optical line emission, not only in the disk plane, but also 
beyond. Other galaxies in our sample exhibit a similar strong 
association of radio continuum and H$\alpha$ emission.

In the following sections we report on our measurements of the
radial extent of both the star-forming parts of the galaxy
disks, $r_{\rm SF}$, and of the associated radio synchrotron 
halos, $r_{\rm halo}$. 
A correspondence between these two radii is a crucial test of 
the dependence of the existence of radio halos on the level of 
SF in the underlying disks.
At the same time we measure the radial extent of the radio 
continuum-emitting disks of the galaxies, $r_{\rm CR}$, and its 
difference compared to the size of the star-forming part, 
$\Delta r = r_{\rm CR} - r_{\rm SF}$.

The measurements are performed by creating slices along the disk
planes of the sample galaxies to determine $r_{\rm CR}$ and
and $r_{\rm SF}$, and additional slices parallel to the disk
to determine $r_{\rm halo}$. The slices have a narrow width
(approximately one beamwidth of our radio continuum images),
over which surface brightnesses are averaged and then plotted
as a function of galactocentric distance. Details on the placement 
and size of the regions are summarised in Table~\ref{tab:cuts}.
The procedures for measuring these quantitites are exactly the 
same as used by us in DLG95. 
In the following subsections we will use data of NGC\,4666 as an
example to demonstrate these techniques.

\subsection{Size of the actively star-forming vs. the radio 
continuum-emitting parts of galaxy disks}

\begin{figure}[b!]
\resizebox{1.0\hsize}{!}{\includegraphics{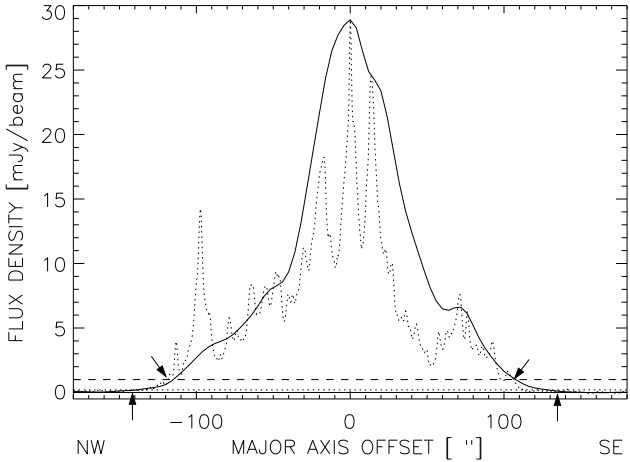}}
\caption{
$12''$ wide cut through our H$\alpha$ image (dotted line) and 
through the 1.49 GHz emission distribution of NGC\,4666 (solid
line) along the disk plane (cf. Fig.~\ref{fig:n4666haz0}), 
demonstrating the measurements of $r_{\rm SF}$ and $r_{\rm CR}$.
Arrows mark the positions at which radii were measured; upwards
arrows at the $r_{\rm CR}$ measurements (at the intersection of the
solid line with the dotted 5-$\sigma$ threshold value), downwards 
arrows at the $r_{\rm SF}$ points (at the intersection of the
dotted data graph with the dashed horizontal threshold line).
}
\label{fig:n4666z0cut}
\end{figure}

In principle, the maximum radii at which SF is observed, $r_{\rm SF}$, 
can be measured from our radio continuum maps, at the radial points 
where the surface brightness of the disk emission drops drastically. 
However, in order to make sure that these measurements are truely 
independent of those of the radial extent of the radio halos 
(below), we choose instead to use H$\alpha$ emission from \hii\ 
regions in the disk as a tracer of SF and thus as a measure of 
$r_{\rm SF}$.
A quantitative measurement can be obtained by setting a threshold 
at a given H$\alpha$ surface brightness, which can be converted 
into a SF  rate. This would require the use of flux-calibrated
H$\alpha$ images, which we do not have available for all targets
at this time.
The placement of the slice through our H$\alpha$ image along the 
disk plane of NGC\,4666 is displayed in Fig.~\ref{fig:n4666haz0}.
Fig.~\ref{fig:n4666z0cut}\ shows the resulting radial profile. Two 
measurements, one on each side of the disk, are obtained, which 
are then averaged. The cutoff criterion for where the star-forming 
disk ends has been set to be on the outer edge of the outermost 
bright \hii-regions in the disk (Figs.~\ref{fig:n4666haz0}\ and
\ref{fig:n4666halocut}).
Although this definition lacks a specific threshold value, the
steepness of the radial brightness profile ensures that the
values for $r_{\rm SF}$ are relatively robust. 
The resulting $r_{\rm SF}$ values are listed in Table~\ref{tab:data}. 

Note that once the correct value of $r_{\rm SF}$ is determined,
one can use integral (IRAS) far-infrared fluxes or radio continuum
data--neither of which are affected by optical thickness--to determine 
average energy input rates by massive SF per unit area, by assuming 
that all emission arises from the star-forming parts of the disks.

\begin{table*}[ht!]
\begin{flushleft}
\leavevmode
\caption{Radii of CR Halos and Comparison with Thin Nonthermal 
  Radio Disks and H$\alpha$ Disk Emission}
\label{tab:data}
\begin{tabular}{lccccccccccc}
\noalign{\hrule\smallskip}
\noalign{\hrule\smallskip}
   ~~~(1)     &  (2)  &  \multispan{2}{~~(3)}{ }  &  
  \multispan{2}{~~(4)}{ }  &  \multispan{2}{~~(5)}{ }  &  (6) 
  & \multispan{2}{~~(7)}{ } & (8) \\
Object     &   $D$  & \multispan{2}{~~$r_{\rm CR}$}{ } & 
  \multispan{2}{~~$r_{\rm halo}$}{ } & \multispan{2}{~~$r_{\rm SF}$}{ }
  & $\Delta r$ & \multispan{2}{~~$r_{25}$}{ } 
  & $A_{\rm SF}\over A_{25}$ \\
           & [Mpc] & [ $''$] & [kpc] &   [ $''$]  & [kpc] &   [ $''$]  & 
  [kpc] & [kpc] & [ $''$]  & [kpc] &  \\
\noalign{\hrule\smallskip}
NGC 891    & 9.5 & 278   & 12.8  &    200  &  9.2 &  203  & 9.3
  & 3.5  & 424 & 19.5 & 0.22\\
NGC 4631 W & 10.0 &  492  & 23.8 &  421 & 20.4 & 362 & 17.6 
  & 6.2  & 444 & 21.5 & 0.67 \\
NGC 4631 E &      &  433  & 21.0 &  338 & 16.4 & 348 & 16.9 
  & 4.1  & 444 & 21.5 & 0.62 \\
\noalign{\hrule\smallskip}
NGC\,1808 & 10.9 &  9$\pm2$ & 0.47$\pm0.10$ & 15$\pm4$ & 0.79$\pm0.21$ 
  &  8$\pm2$ & 0.42$\pm0.10$ & 0.05$\pm0.15$ & 195 & 10.3 & 0.0017 \\
M\,82    &  3.2 & 46$\pm3$ & 0.71$\pm0.03$ & 60$\pm3$ & 0.93$\pm0.05$ 
  & 21$\pm2$ & 0.33$\pm0.04$ & 0.38$\pm0.06$ & 336 & 5.2 & 0.0039 \\
NGC\,4666 & 26.4 & 119$\pm4$ & 15.2$\pm0.5$ & 90$\pm6$ & 11.5$\pm0.8$ 
  & 106$\pm4$ & 13.6$\pm0.5$ & 1.6$\pm0.7$ & 138 & 17.7 & 0.59 \\
NGC\,4700 & 25.5 & 63$\pm5$ & 7.7$\pm0.5$ & 42$\pm7$ & 5.2$\pm0.6$ 
  & 51$\pm7$ & 6.3$\pm0.5$ & 1.1$\pm0.6$ & 90 & 11.1 & 0.35 \\
NGC\,7090 & 11.7 & 147$\pm8$ & 8.3$\pm0.5$ & 93$\pm12$ & 5.3$\pm0.7$
  & 74$\pm8$ & 4.2$\pm0.5$ & 0.9$\pm0.6$ & 222 & 12.6 & 0.35 \\
\noalign{\smallskip\hrule}
\end{tabular}
\end{flushleft}
The entries in  Table~\protect\ref{tab:data} are: \\
(1) Galaxy name; (2) Distances are based on $H_0$ = 75 km s$^{-1}$ 
Mpc$^{-1}$ and a virgocentric infall velocity of 300 km s$^{-1}$ (cf. 
Dahlem et al. 2001); (3) Radial extent of the radio continuum-emitting 
disk; (4) Radial extent of the radio synchrotron halo; (5) Radial 
extent of the star-forming disk, measured from H$\alpha$; (6) 
Difference between $r_{CR}$ and $r_{SF}$; (7) Radius of stellar 
disk at 25th magnitude (in the B-band); (8) Scaling factor between 
star-forming part of the disk, $A_{\rm SF}$, vs. stellar disk extent, 
$A_{25}$.
\end{table*}

The values for the radii of the radio continuum-emitting disks, 
$r_{\rm CR}$, were measured at the 5-$\sigma$ confidence level of 
the images by producing cuts through the radio continuum emission 
distribution in the galaxy disks along their major axes, as 
described by us in DLG95. Again, two measurements are taken and 
then averaged.
The placement of the cut through the 1.49 GHz disk emission of 
NGC\,4666 is exactly the same as that for the H$\alpha$ cut
(Fig.~\ref{fig:n4666haz0}). The resulting surface brightness
profile as a function of galactocentric distance is shown in 
Fig.~\ref{fig:n4666z0cut}\ as a solid line. Details are listed
in Table~\ref{tab:cuts}\ and the resulting measurements of
$r_{\rm CR}$ can be found in Table~\ref{tab:data}.

\begin{figure}[b!]
\resizebox{1.0\hsize}{!}{\includegraphics{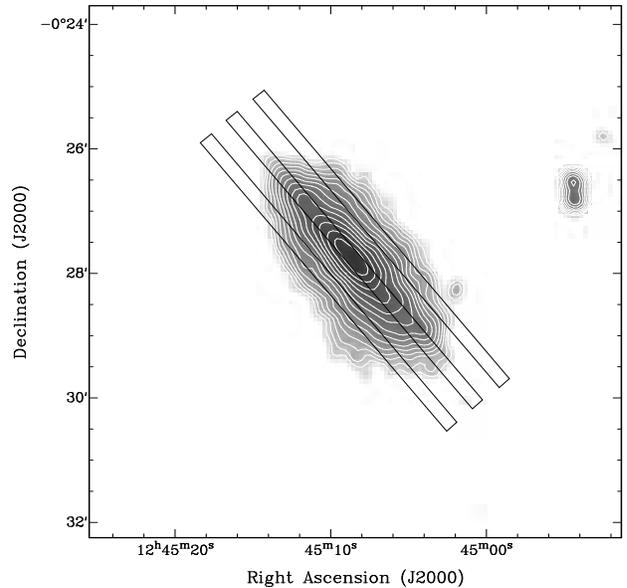}}
\caption{
Overlay of rectangular boxes with widths of $12''$ for the
measurements of $r_{\rm CR}$ and $r_{\rm halo}$ in NGC\,4666 on our
1.49 GHz radio continuum image. The central cut is the same as in
Fig.~\ref{fig:n4666haz0}, the properties of the two parallel cuts
through the halo are listed in Table~\protect\ref{tab:cuts}.
}
\label{fig:n4666cuts}
\end{figure}

\subsection{Radial extent of the radio halos, $r_{\rm halo}$}

\begin{figure}
\resizebox{1.0\hsize}{!}{\includegraphics{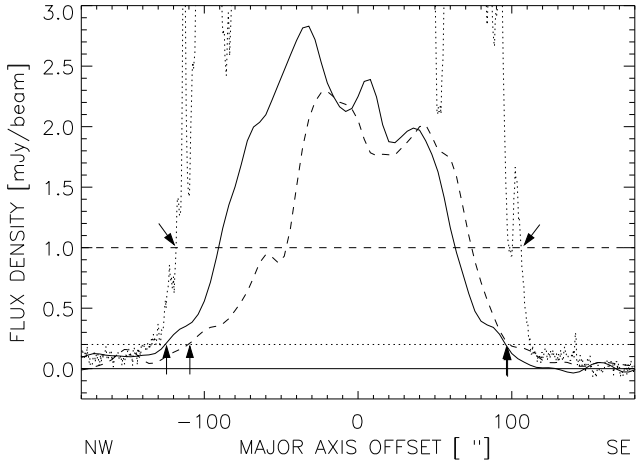}}
\caption{
Cuts through the 1.49 GHz emission distribution of NGC\,4666 
through the radio halo, parallel to the disk plane, at $z$-offsets of
$\pm32''$ (see Table~\ref{tab:cuts}). 
The cut through the halo north-west of the plane is shown as a dashed
line, the south-eastern cut is represented by a solid line.
For comparison the cut through the H$\alpha$ image along the disk plane
is shown again as a dotted line. Arrows mark the positions at which 
radii were measured; upwards arrows at the $r_{\rm halo}$ measurements, 
downwards arrows at the $r_{\rm SF}$ points (same as in 
Fig.~\protect\ref{fig:n4666z0cut}).
}
\label{fig:n4666halocut}
\end{figure}

We measure the radial extent of the radio halos, $r_{\rm halo}$, 
by producing cuts through the radio continuum emission distribution 
in the halos, parallel to the galaxies' major axes, but offset from 
the disk planes (DLG95). Measurements are again taken at the 
5-$\sigma$ confidence level. 
We show the positions of the cuts through the radio halo of NGC\,4666 
in Fig.~\ref{fig:n4666cuts}, where we follow exactly the same procedure 
as in DLG95 (cf. Fig.~3 in that paper). The placement of the cuts is
quantified in Table~\ref{tab:cuts}.
From the two cuts through the halo of each galaxy, one can ideally 
obtain four measurements of $r_{\rm halo}$. Because axial symmetry 
is a good assumption for the relatively undisturbed systems studied 
here, the four values can be averaged into one per galaxy. 
Fig.~\ref{fig:n4666halocut}\ displays, as an example, the points
at which $r_{\rm SF}$ and $r_{\rm halo}$ were measured in NGC\,4666.
In practice, some values cannot be used, e.g. because of the presence
of nearby background sources. Column~6 in Table~\ref{tab:cuts}\
provides the number of measurements used to obtain $r_{\rm halo}$ for 
each galaxy.
The average values for $r_{\rm halo}$ obtained by us (in angular and
spatial units) are presented in column~4 of Tab.~\ref{tab:data}.
Since it is important that the cuts in the halo not be contaminated 
by disk emission (by either beam-smearing, a warped disk or a 
deviation of inclination from edge-on [i=90$^\circ$]), $z$-offsets
where chosen to be more than one beam-width. At the same time, the
$z$-offsets must not be too large because the signal-to-noise ratio
of the data drops rapidly (exponentially) away from the galaxy disks.
Accordingly, cuts were produced at intermediate $z$-offsets, well
above the projected disks of the galaxies, but as low as possible
in order to achieve the highest possible signal-to-noise ratios.
We list the $z$-offsets of the cuts from the major axes in units of 
arcsec, kpc and beamwidths in Table~\ref{tab:cuts}.

All average values for the three radii ($r_{\rm SF}$, $r_{\rm CR}$
and $r_{\rm halo}$) are collated in Table~\ref{tab:data}. The values 
in the top part of the table, separated by a horizontal line, are 
reproduced from DLG95, the lower part of the table contains new 
measurements.

Our estimates of the uncertainties for the various radii are dominated
by uncertainties arising from the limited angular resolution of the
radio images. In addition, the observed radio halos are not exactly 
cylindrical and therefore there is a weak dependence of the radial 
extent, $r_{\rm halo}$, on the $z$-offset at which the measurement 
is obtained. Because of the constraints mentioned before, this 
uncertainty can at present not be avoided.

Due to signal-to-noise limitations, studies of variations of the halo 
width as a function of $z$-distance can currently not be conducted yet.

\section{Discussion}
\label{par:disc}

Our new results enable us to perform the same studies on a sample 
of five new galaxies as done previously for NGC\,891 and NGC\,4631 
(DLG95), with which they can be compared:

\begin{enumerate}
\item Check whether there is a relation between the maximum radius
      at which massive SF is observed, $r_{\rm SF}$, and the radial 
      extent of the radio halo, $r_{\rm halo}$ (Sec.~\ref{par:radext});
\item calculate the {\it average} energy input rate from high-mass SF 
      in the disk, \edota, and determine the threshold value for the 
      onset of halo emission (Sec.~\ref{par:einput}).
\end{enumerate}

In addition, the slowly increasing size of our sample now also allows 
us to investigate whether the existence and properties of gaseous 
halos depend on a number of other global physical galaxy properties 
(Sec.~\ref{par:halodep}). For this study values obtained for other 
galaxies (DLG95) have been reused.

\begin{figure*}
\resizebox{0.5\hsize}{!}{\includegraphics{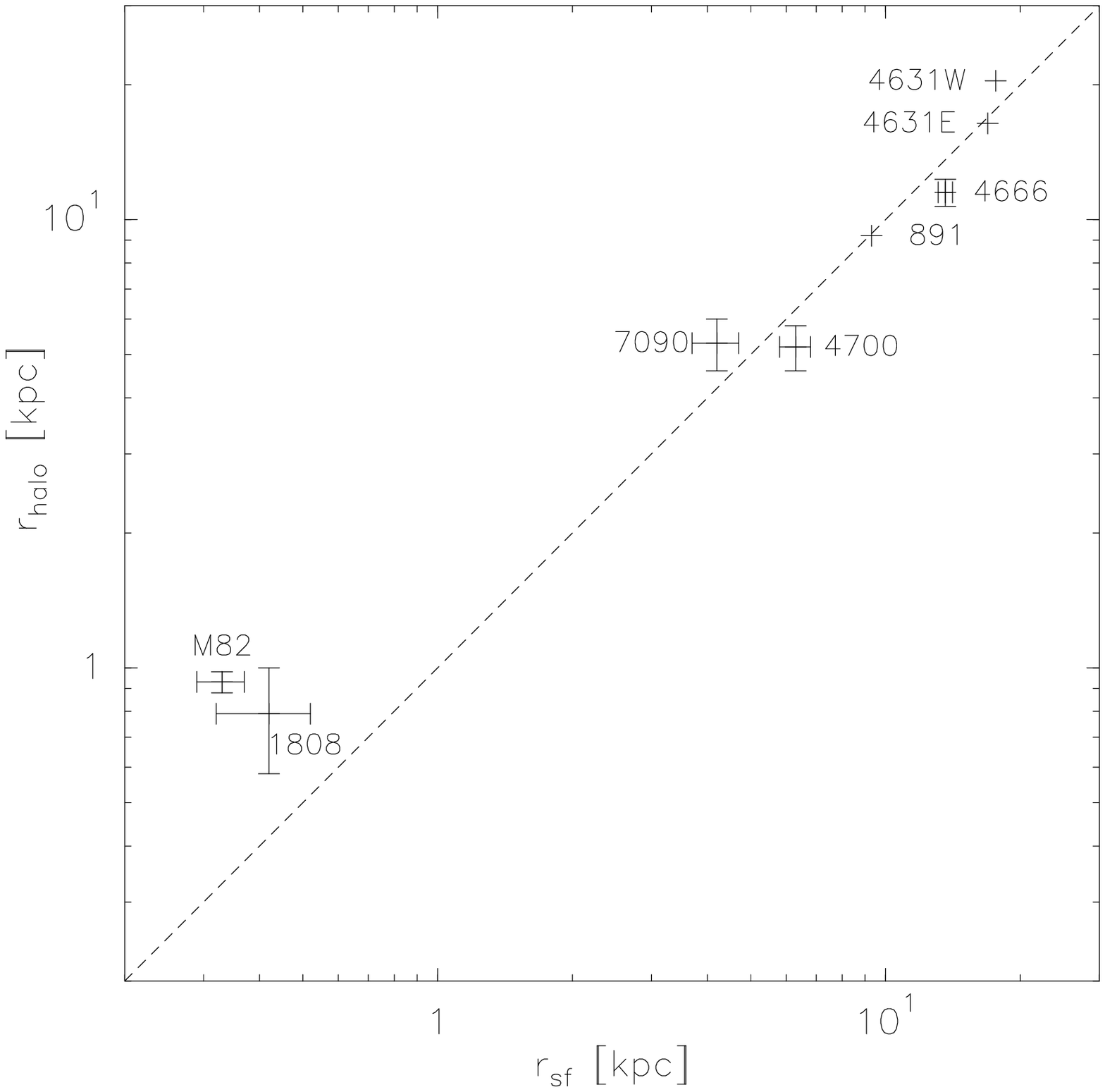}}
\resizebox{0.5\hsize}{!}{\includegraphics{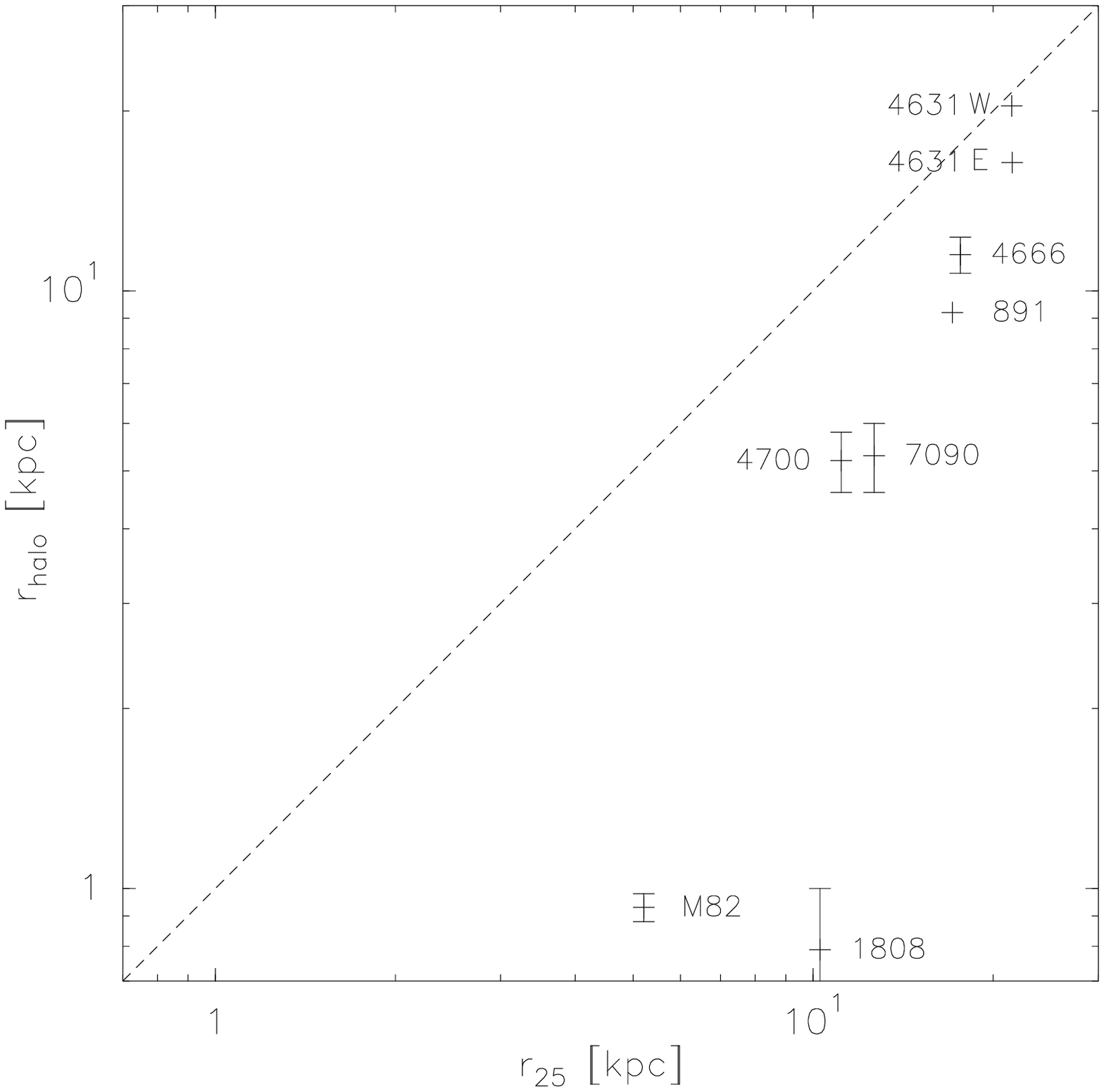}}
\caption{
{\bf Left:} Radial extent of the star-forming part of the galaxy disks,
$r_{\rm SF}$, against the radial extent of the associated 
halos, $r_{\rm halo}$; see Sec.~\protect\ref{par:radext}\ for details.
{\bf Right:} Radial extent of the optical disk, $r_{25}$,
against $r_{\rm halo}$.
}
\label{fig:rhalovsrsf}
\end{figure*}

\subsection{Star-forming disk vs. radial halo extent}
\label{par:radext}

As before (DLG95), we find a strong dependence of the halo 
properties on the level of disk activity in our sample
galaxies.
The numbers quantifying these relationships are collated
in Table~\ref{tab:data}. 
A tight and linear correlation between the radial extent 
of radio halos, $r_{\rm halo}$, and the areas of active SF in 
the underlying disks, $r_{\rm SF}$, is confirmed. This is 
visualised in the left panel of Fig.~\ref{fig:rhalovsrsf}. 
A weaker and strongly nonlinear dependence is visible when 
plotting $r_{\rm halo}$ against the radial extent of the optical 
disk at the 25th magnitude surface brightness, $r_{25}$ (right 
panel of Fig.~\ref{fig:rhalovsrsf}).
The use of $r_{25}$ is not meaningful in this context, as can 
be judged by considering a study of galaxies with proto-typical
circumnuclear starbursts. In these galaxies, the vast difference 
between $r_{25}$ and $r_{\rm SF}$ (see entries for NGC\,1808 and 
M\,82 in Table~\ref{tab:cuts}) illustrates clearly that $r_{25}$
is not a suitable tracer of currently ongoing SF activity. This,
in turn, supports the notion that the observed gaseous halos
are created by processes related to the young stellar populations
(I), not the old ones (II).

The extent of the radio halos, $r_{\rm halo}$, of galaxies with 
widespread SF is similar to that of the underlying star-forming 
disks, $r_{\rm SF}$, because in the outer parts of the galaxy
disks the SF rates (SFRs) and thereby energy input rates
are too low to create outflows (cf. DLG95). 
The $r_{\rm halo}$ values are in most cases slightly smaller than 
$r_{\rm SF}$, which is probably a sensitivity bias. These data 
points lie just below the line of unity in the left panel of
Fig.~\ref{fig:rhalovsrsf}.

A direct comparison of the values of $r_{\rm SF}$ and $r_{\rm 
halo}$ (Table~\ref{par:data}) suggests that for circumnuclear 
starbursts (NGC\,1808, M\,82) $r_{\rm halo}$ is slightly {\it 
larger} than $r_{\rm SF}$. 
This is caused by the fact that the starburst-driven outflow 
cones widen beyond the disk planes in the arche-typical 
hour-glass shape (e.g. Veilleux et al. 2005 and references 
therein). For this reason, classical starbursts will always
appear slightly above the line of unity in Fig.~\ref{fig:rhalovsrsf}
(left panel).

The significance of a direct relation of the halo extent and 
the maximum radius at which a galaxy is actively forming stars
is that--based on the data used here--halos appear to exist
only when and where the current massive SF rate is high. It 
corroborates calculations predicting outflows from active SF 
regions predominantly perpendicular to the disk plane due to 
the pressure gradient along which material can flow. At the 
same time expansion within the disk is inhibited by the
resistance of swept-up disk material.

One example of a model making such predictions, although too
simplistic to explain all details, is the ``chimney'' model 
by Norman \& Ikeuchi (1989). Any modern model of the interstellar
medium in galaxies contains a halo component, which is created
naturally as a consequence of the buoyancy of disk gas heated
by massive SF (see, e.g,. Mac Low \& Ferrara 1999). 
Bregman (1980) pointed out that the conservation of angular
momentum during the transport of matter into a galaxy's halo
(in the form of a ``galactic fountain'') leads to radial motions, 
thus facilitating an efficient radial re-distribution of 
metal-enriched material within galaxies.

The major advance of our current observational results over 
DLG95 is that our findings are now firmer, because based on 
significantly more sensitive observations of more objects 
(seven compared to two) than before.

\subsection{Extent of the star-forming disk vs. total radio 
continuum-emitting disk}

There are two reasons why a significant difference between the 
extent of an actively star-forming disk and that of the radio
continuum-emitting part of a galaxy can be observed: \\
1. Diffusion of CRs from sites of massive SF. In the case of
   radial diffusion within the disk, for galaxies in which the
   level of SF drops rapidly at a certain galactocentric radius,
   $\Delta r = r_{\rm CR}\ - r_{\rm SF}$ is a measure of the
   radial CR diffusion coefficient within the disk, $D$ (DLG95). \\
2. The presence of low-level SF in the outermost parts of the 
   disks can lead to low surface brightness radio emission. 

Galaxies with widespread SF (in the present sample NGC\,4666, 
NGC\,4700 and NGC\,7090; NGC\,891 and NGC\,4631 from DLG95) 
exhibit a significant difference between $r_{\rm SF}$ and 
$r_{\rm CR}$ (see Table~\ref{tab:data}). This suggests either 
radial electron diffusion in the disks or low-level SF in the 
outer parts of their disks, or a combination of both.

On the other hand, the measured differences between $r_{\rm SF}$ 
and $r_{\rm CR}$ within the disks of circumnuclear starbursts are 
insignificant. 
A likely cause of this similarity could be that radial propagation 
of CRs within the disks is inhibited by the high pressure of the 
ambient medium and possibly by poloidal magnetic fields, which are 
found in powerful starburst galaxies (e.g., Lesch et al. 1990).
At the same time, there is little or no SF in the disks of some
circumnuclear starburst galaxies beyond their Inner Lindblad
Resonance (or turnover radius of rotation); see e.g. Combes (1987),
Lesch et al. (1990).

\subsection{Energy input rates into the disk ISM}
\label{par:einput}

\begin{table*}[t!]
\begin{flushleft}
\leavevmode
\caption{Energy Input by Supernovae Derived from Radio Data}
\label{tab:e_sn}
\begin{tabular}{lcrrcccc}
\noalign{\hrule\smallskip}
\noalign{\hrule\smallskip}
 ~~~(1)   &   (2)   &   (3)~   &   (4)~   &   (5)   &   (6)  &  (7)  \\
~Object     & Morph. & $D$~~ & $r_{SF}$ & $P_{\rm nth}$ & 
  $\nu_{SN}$  &  \edota \\
 & Type$^{RC3}$ & [Mpc]  & [kpc] & [$10^{21}$ W\,Hz$^{-1}$] &
  [yr$^{-1}$]  &  [10$^{-3}$ \ergscm]  \\
\noalign{\hrule\smallskip}
NGC\,\,~891  & SA(s)b? sp  & 9.5     &  9.3 & 8.2   & 0.13 & 1.60 \\
NGC\,3044   & SB(s)c? sp  & 20.6    & 10.0 & 6.1   & 0.10 & 1.02 \\
NGC\,4631   & SB(s)d      & 10.0    & 17.3$^{\rm a}$ & 14.7~  & 0.23 & 0.82 \\
NGC\,5907   & SA(s)c: sp  & 14.9    & 18.1 & 2.6   & 0.04 & 0.13 \\
NGC\,4565   & Sbc         &  9.7    & 16.9 & 1.5   & 0.02 & 0.09 \\
NGC\,4244   & SA(s)cd? sp &  3.1    & 4.5  & 0.1  & 0.002 & $<0.08$~~~ \\
\noalign{\hrule\smallskip}
M\,82      & I0;Sbrst \hii &  3.2 & 0.33 & 10.1~ & 0.16 & 1550~ \\
NGC\,1808  & (R'\_1)SAB(s:)b  & 10.9 & 0.42 & 5.9 & 0.09 & 565 \\
NGC\,4666  & SABc:       & 26.4 & 13.7 & 34.1   & 0.53 & ~3.1 \\
NGC\,3175  & SAB(s)b & 15.9 & 3.5  & ~~2.1$^+$  & 0.03 & 2.98 \\
NGC\,4700  & SB(s)c? sp; \hii  & 25.5 & 6.3 & 2.3 & 0.04 & 0.99 \\
NGC\,1406  & SB(s)bc: sp & 14.9 & 7.9  & ~~3.4$^+$  & 0.05 & 0.90 \\
NGC\,3437  & SAB(rs)c: & 25.5 & 11.8 & ~~5.7$^+$  & 0.09 & 0.69 \\
NGC\,7090  & SBc? sp & 11.7 & 4.2 & ~~0.6$^*$  & 0.01 & 0.59\\
NGC\,2748  & SAbc \hii & 28.7 & 14.3 & ~~6.6$^+$  & 0.10 & 0.54 \\
NGC\,1421  & SAB(rs)bc: & 31.1 & 21.8 & ~12.8$^+$  & 0.20 & 0.45 \\
NGC\,7462  & SB(s)bc? sp & 15.1 & 5.4  & ~~0.7$^+$  & 0.01 & 0.41 \\
NGC\,1055  & SBb: sp LINER2 & 16.0 & 18.3 & ~~7.2$^+$  & 0.11 & 0.36 \\
NGC\,3717  & SAb: sp \hii & 27.1 & 23.2 & ~10.8$^+$ & 0.17 & 0.34 \\
NGC\,5297  & SAB(s)c: sp & 34.5 & 12.7 & 3.3 & 0.05 & 0.34 \\
NGC\,2613  & SA(s)b & 25.9 & 21.2 & 5.5 & 0.08 & 0.21 \\
NGC\,4517  & SA(s)cd: sp & 19.5 & 18.4 & 1.7 & 0.03 & 0.08 \\
\noalign{\smallskip\hrule}
\end{tabular}
\end{flushleft}
The entries in  Table~\protect\ref{tab:e_sn} are: \\
(1) Galaxy name; (2) Morphological type from NED, based on the RC3 
catalogue; (3) Distances are based on $H_0$ = 75 km s$^{-1}$ Mpc$^{-1}$ 
and a virgocentric infall velocity of 300 km s$^{-1}$ (cf. Dahlem et
al. 2001); (4) Radial extent of the star-forming disk, measured from 
H$\alpha$ imagery; (5) Total nonthermal radio power at 1.49 GHz (1.384 
GHz for galaxies marked with an asterisk [$^*$]; 1.43 GHz for galaxies 
marked with a cross [$^+$]); (6) Supernova rate derived from the total 
radio power, $P_{\rm nth}$, applying eq.~4 of DLG95; (7) The energy 
input per SF area derived from eq.~8 of DLG95, assuming for all galaxies 
a magnetic field strength of 5 $\mu$G. \\
Note to Table~\protect\ref{tab:e_sn}: \\
a) Average value for both sides of the disk
\end{table*}

Based on $r_{\rm SF}$ we calculate the surface area within the 
sample galaxies over which SF occurs, assuming circular symmetry, 
$A_{\rm SF} = \pi\ r_{\rm SF}^2$. This we adopt as the surface 
area over which energy is released homogeneously into the ambient 
medium by SF-related process, i.e. stellar winds, supernovae and 
their remnants (Leitherer and Heckman 1995), \esn.
The energy input is assumed to be proportional to the nonthermal
radio emission and its calculation and normalisation to unit 
surface area, \edota\ = \esn/$A_{\rm SF}$, are performed exactly 
as described in DLG95. 
The relevant numbers are collated in Tables~\ref{tab:data}\ and
\ref{tab:e_sn}, where the top parts contain values from DLG95, 
while the lower parts add new measurements to the database.

Especially for the classical circumnuclear starbursts the use of 
$r_{\rm SF}$ instead of $r_{25}$ is obviously essential. 
Neglecting this correction would lead to errors in estimates 
of the mean energy input rates per unit surface area, \edota, by 
factors of 1000, typically, as can be seen by inverting the numbers 
in column 8 of Table~\ref{tab:data}.
But also for galaxies with widespread SF corrections in \edota\ 
by factors of 1.5--5 occur when using $r_{\rm SF}$ instead of 
$r_{\rm 25}$ (Tables~\ref{tab:data}\ and \ref{tab:e_sn}).

%
%
Again, there is a clear difference between proto-typical starburst 
galaxies (NGC\,1808, M\,82) and others with more widespread SF. 
The energy injection rates in circumnuclear starbursts, \edota, are 
orders of magnitude higher than in galaxies with widespread SF, 
leading to much more energetic outflows which can be detected at 
high distances above the disk planes, of up to 5--10 kpc (Dahlem 
et al. 1998, Lehnert et al. 1999, Strickland et al. 2004a,b, 
T\"ullmann et al. 2006). In these outflows gas can reach of order 
escape velocity (e.g., HAM90) and shock heating plays an important 
role (e.g., Chevalier and Clegg 1985, HAM90, Dahlem et al. 1997).

%
%
Outflows from galaxies with widespread SF reach lower maximum 
$z$-heights of up to 1--5 kpc (Rand 1996, Rossa \& Dettmar 
2000, 2003a,b, Miller \& Veilleux 2003), because the average energy 
input rates per unit surface area are lower than in starbursts,
although the total energy input may be similar (see column 5 in
Table~\ref{tab:e_sn}).

\subsection{Halo dependence on various physical properties}
\label{par:halodep}

Several quantities were investigated for potential relationships.
In particular, possible dependences of the halo properties on
the level of SF in the galaxy disks, as measured by the energy
deposition rate \edota, and on galaxy mass surface density
were tested.

\begin{table*}[t!]
\begin{flushleft}
\leavevmode
\caption{Fundamental Properties of Galaxies Shown in 
  Figures~\protect\ref{fig:edotavsmass}--\protect\ref{fig:sfvsfircol}}
\label{tab:fir}
\begin{tabular}{lrrcccccc}
\noalign{\hrule\smallskip}
\noalign{\hrule\smallskip}
   ~~~(1)     &  (2)~  &  (3)  &   (4)  &    (5)  &   (6)  &  (7)  
  &  (8)  &  (9)  \\
Object     &  $f_{60}$~ & $f_{100}$ & ${f_{60}\over f_{100}}$ & \lfir 
  & \lb & m$_{\rm K}$ & \mk & \mk/A$_{\rm 25}$ \\ 
           &   [Jy]  &  [Jy]  &  & [$10^{10}$ \lsol] & [$10^{10}$ \lsol] 
  & [mag] &  [$10^{10}$ \msol] &[\msol\ pc$^{-2}$] \\
\noalign{\hrule\smallskip}
NGC\,~891 &   61.10 &  198.63 & 0.31 & 1.27 & 1.44 & 5.94 & 3.2 & 27.0\\
NGC\,1055 & 23.27 &  60.09 & 0.39 & 1.21 & 1.18 & 7.15 & 3.0 & 30.6\\
NGC\,1406 & 11.91 &  27.00 & 0.44 & 0.51 & 0.54 & 8.61 & 0.7 & 31.9\\
NGC\,1421 & 11.20 &  24.30 & 0.46 & 2.03 & 3.43 & 8.40 & 3.6 & 44.5\\
NGC\,1808 & 87.81$^{\rm a}$ & 137.20$^{\rm a}$ & ~0.64$^{\rm a}$ 
  & ~1.70$^{\rm a}$ & 0.70 & 6.66 & 2.2  & 65.9\\
NGC\,2613 & 6.31  &  25.42 & 0.25 & 1.10 & 6.37 & 6.83 & 10.6~  & 47.6\\
NGC\,2748 &  7.95 &  19.44 & 0.41 & 1.30 & 1.70 & 10.32~ & 0.5 & 10.6\\
M\,82     & 1271.32 & 1351.09 & 0.94 & 1.87 & 0.26 & 4.67 & 1.2  & 138.9~~ \\
NGC\,3044 &   10.47 &   21.16 & 0.49 & 0.81 & 1.34 & 8.98 & 0.9 & 15.6\\
NGC\,3175 & 13.07 &  28.20 & 0.46 & 0.62 & 0.73 & 7.79 & 1.6  & 61.6\\
NGC\,3437 & 12.15 &  20.62 & 0.59 & 1.33 & 0.92 & 8.88 & 1.5 & 57.6\\
NGC\,3717 & 10.52 &  23.84 & 0.44 & 1.48 & 2.67 & 7.52 & 6.1 & 35.0\\
NGC\,4244 &    4.20 &   16.06 & 0.26 & 0.01 & 0.17 & 7.72 & 0.07 & ~4.8\\
NGC\,4517 & 6.92  &  20.20 & 0.34 & 0.57 & 3.61 & 7.33 & 3.8  & 14.3\\
NGC\,4565 &    9.83 &   47.23 & 0.21 & 0.27 & 1.92 & 6.06 & 3.0 & 22.0\\
NGC\,4631 &   82.90 &  208.66 & 0.40 & 1.67 & 3.21 & 6.67 & 2.2  & 15.2\\
NGC\,4666 &   37.34 &   82.88 & 0.45 & 4.92 & 3.32 & 7.06 & 8.9 & 91.1\\
NGC\,4700 &    3.05 &    5.36 & 0.57 & 0.34 & 1.54 & 9.78 & 0.7 & 17.4\\
NGC\,5297 & 1.81  &  7.99  & 0.23 & 0.59 & 2.95 & 9.89 & 1.1  & ~5.4\\
NGC\,5907 &   12.02 &   42.37 & 0.28 & 0.64 & 2.61 & 6.76 & 3.7 & 18.4\\
NGC\,7090 &    5.88 &   17.97 & 0.33 & 0.18 & 0.97 & 8.16 & 0.6 & 12.7\\
NGC\,7462 & 3.11 &  6.32 & 0.49 & 0.13 & 0.55 & 9.65 & 0.3  & 10.0\\
\noalign{\smallskip\hrule}
\end{tabular}
\end{flushleft}
The entries in  Table~\protect\ref{tab:fir} are: \\
(1) Galaxy name; (2) IRAS 60 $\mu$m band flux density; from NED; (3) IRAS 
100 $\mu$m band flux density; from NED; (4) $f_{60}/f_{100}$ flux ratio;
(5) Far-infrared luminosity, based on the standard formulas used by us
earlier (Dahlem et al. 2001); (6) Optical blue luminosity, determined as 
\lb = $89.13\times10^{-0.4\times M_{\rm B}}$, where the absolute blue 
magnitude is the apparent blue magnitude, K-corrected, corrected for 
Galactic extinction and scaled by the distance of the object, $M_{\rm B} 
= m_{\rm B,0}^{\rm T}$-25-5$\times$log($D$); $m_{\rm B,0}^{\rm T}$ values 
are from NED. This definition provides an estimate of the blue luminosity 
($\nu L_\nu$) at 4400 \AA; (7) Apparent total magnitude in the K-band 
(centered at 2.17 $\mu$m) from 2MASS; (8) Mass in galactic disk derived 
from the K-band magnitude, as described in Sec.~\ref{par:edotavsmass};
(9) Average mass surface density within the optical (r$_{25}$) disk. \\
Note to Table~\protect\ref{tab:fir}: \\
a) Not used for studies of FIR properties due to unknown contribution
   from a low-luminosity AGN.
%
%
\end{table*}

\subsubsection{Energy injection rate vs. galaxy mass surface density}
\label{par:edotavsmass}

Lower-mass galaxies not only have lower escape velocities 
than more massive ones, but the onset of disk-halo interactions
will also be easier for a shallower gravitational potential.
Therefore, the existence of a radio halo is expected to depend, 
apart from the energy input from the disk (see above), on the 
gravitational force that particles must overcome to escape from 
a galaxy disk. 
The gravitational potential that particles experience when being 
lifted above the disk is proportional to the mass surface density 
in the disk. We approximate the mass in the disk by the stellar 
mass derived from near-infrared emission, which traces old stars
that are responsible for the bulk of the stellar mass. 

The stellar mass is deduced from the total K-band magnitude, 
$m_{\rm K}$, from the Two Micron All Sky Survey (2MASS). The 
luminosity in this band, $L_{\rm K}$, was calculated as

\begin{equation}
L_{\rm K} = 2.8 \times 10^{10} \frac{D^2}{[\rm Mpc]} 
10^{-0.4 m_{\rm K}}\ [{\rm L_\odot}],
\end{equation}
which gives an estimate of the luminosity ($\nu\ L_\nu$) at 2.17
$\mu$m.

From the luminosity the stellar mass was derived assuming 
optically thin emission (which may lead to a slight underestimate 
of the NIR emissivity) and a mass-to-light ratio of 3, which is 
indicative of a stellar population of age 5.9 Gyr (Bruzual \& 
Charlot 2003).
The K-band magnitudes and \mk\ values for the galaxies in our sample 
are listed, together with other parameters, in Table~\ref{tab:fir}.

\begin{figure*}
\resizebox{0.5\hsize}{!}{\includegraphics{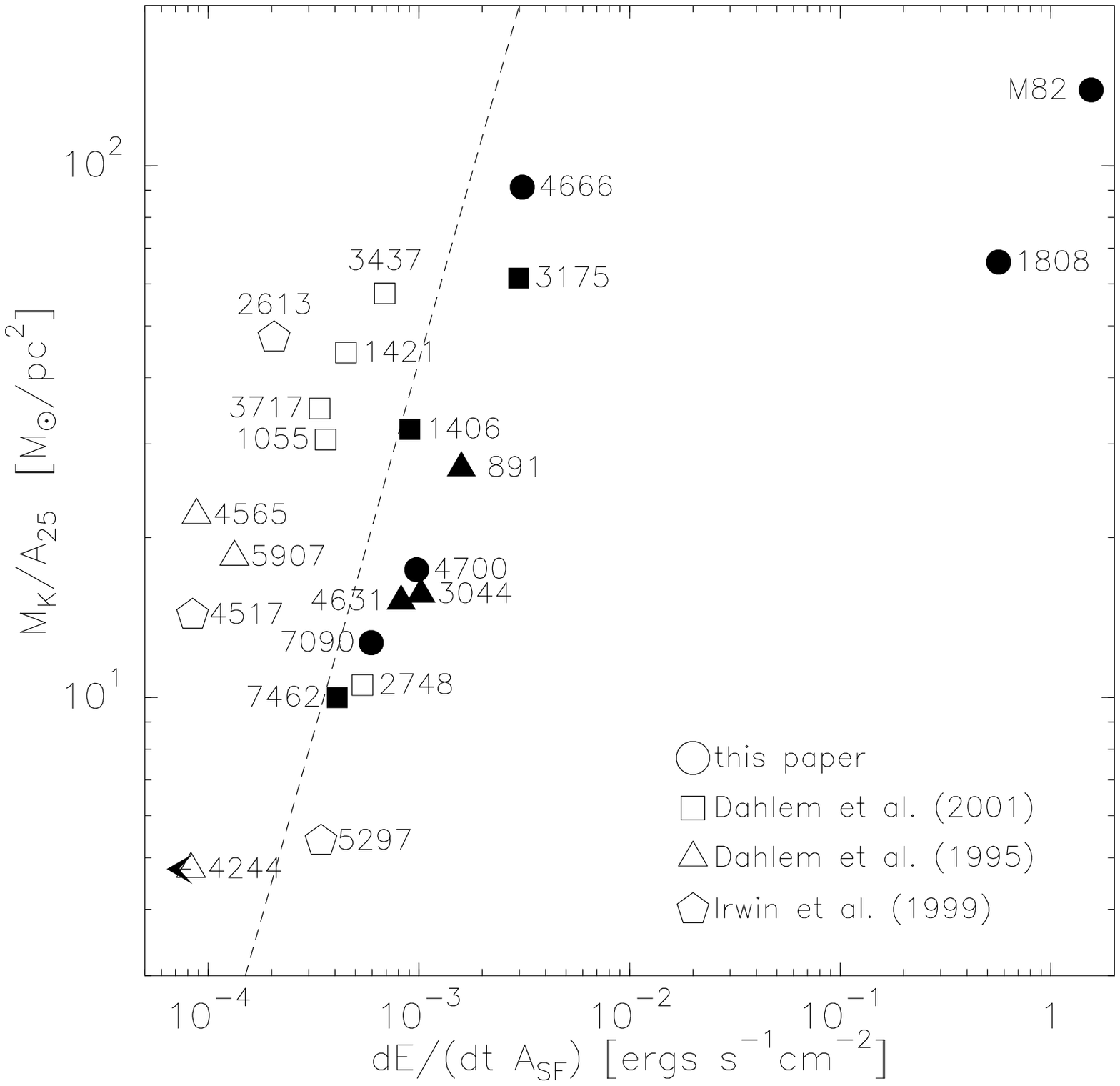}}
\resizebox{0.5\hsize}{!}{\includegraphics{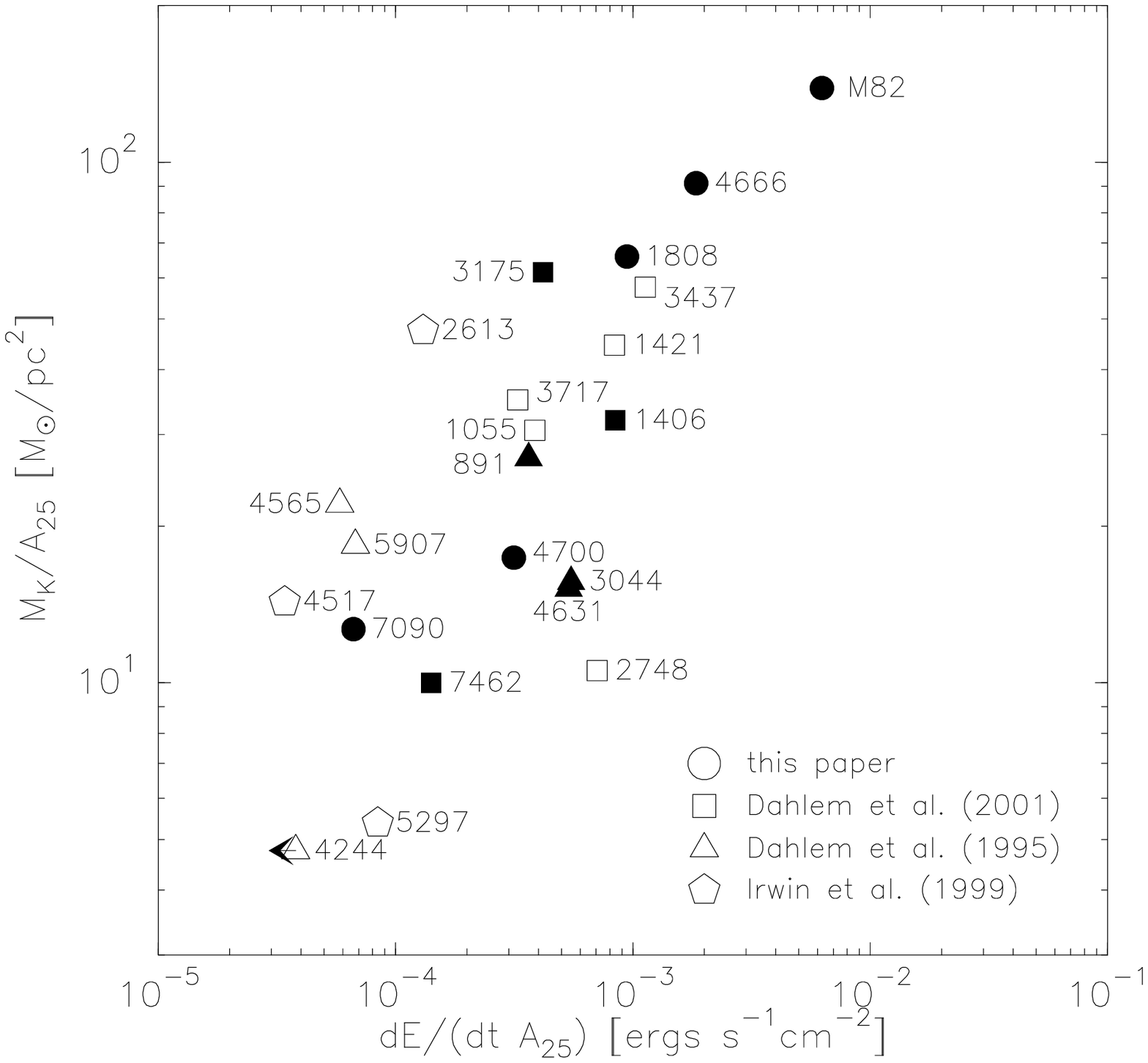}}
\caption{
{\bf Left:} Energy input per unit star-forming area, $A_{\rm SF}$, 
as a function of stellar mass surface density, as derived from K-band 
imagery. Filled symbols indicate galaxies with radio halos, open symbols 
are galaxies without radio halo detections. The dashed line roughly 
indicates the transition zone between galaxies with and without radio 
halos.
{\bf Right:} The same as on the left, but using $A_{25}$ instead of 
$A_{\rm SF}$ as an approximation of the area over which energy is 
deposited into the disk ISM.
}
\label{fig:edotavsmass}
\end{figure*}

The significance of the rate at which energy is deposited into 
the ISM of spiral galaxies, \edota, on the initiation of disk-halo 
outflows, and thereby the creation of gaseous halos, has been 
described above (Sec.~\ref{par:einput}). Only galaxies with energy 
input rates above a certain threshold value can start disk-halo 
interactions. Note that \edota\ values can be converted directly 
into the commonly used, but more indirect, SF rates per unit area, 
which are often measured in units of \msol\ yr$^{-1}$ kpc$^{-2}$.

In order to investigate the influence of both the galaxy mass surface  
density and the energy input rates, 
we plot in the left panel of Fig.~\ref{fig:edotavsmass}\ the stellar
surface density, \mk/$A_{25}$, where $A_{25} = \pi r_{25}^2$ is the 
area of the optical disk, versus \edota\ for the galaxies in our 
extended sample. This plot is similar to the one shown in Lisenfeld 
et al. (2004), however now using the stellar mass in the disk, rather 
than the total dynamical mass. In addition, the present plot also 
includes more new galaxies from the present paper.

As expected, we note a clear division between galaxies with and 
without radio halo: 
Galaxies with low mass surface densities and high energy input rates 
have radio halos, as opposed to galaxies with high mass surface 
densities and low energy  input rates. The line drawn indicates the 
transition between both regimes. 

The significance of the observed dichotomy is that, because of the 
smaller force that particles must overcome and the associated lower 
escape velocities, it is easier for low-mass galaxies to have gaseous 
halos around them than massive ones (e.g. Heckman et al. 1995, Martin 
1997, 1998).
Once the energy input rate is high enough, also spiral galaxies with 
a high mass surface density can create and maintain gaseous halos.
The energy input levels in classical starburst galaxies are shown 
to be so high (Fig.~\ref{fig:edotavsmass}, left panel) that 
gravitational forces are negligible, as already assumed by Chevalier 
and Clegg (1985).

These new findings refine our statement (DLG95) that there is ``a 
threshold energy input rate'' above which outflows can occur 
by including the galaxy mass surface density as an additional 
parameter.
The energy threshold value determined here for galaxies of given 
mass surface density for the onset of disk-halo interactions can be
used in the future to predict whether or not a certain galaxy can be 
expected to have a gaseous halo, i.e. whether the ``breakout 
condition'', as defined by Norman \& Ikeuchi (1989), required 
for outflows to occur is fulfilled.
The detected \edota\ vs. \mk/$A_{25}$ relation is reliable, 
because it does not depend strongly on the adopted galaxy distances, 
both quantities essentially being derived from surface brightnesses.

In this context it is important to note that using $r_{25}$ instead 
of $r_{\rm SF}$ to measure \edota\ would lead to the displacement of 
objects along the x-axis to lower values of \edota, which would skew 
the detected \edota\ vs. \mk/($A_{25}$) relationship. This is
illustrated in the right panel of Fig.~\ref{fig:edotavsmass}.
Accordingly, no such clear separation between galaxies with and
without a halo is visible there and the placement of data points
for classical starbursts is completely wrong.

Strongly interacting galaxies must be excluded from studies of the 
dependence of halo properties on the underlying SF activity, because 
of the potential influence of the gravitational force of a companion 
galaxy on the gas kinematics and thus disk-halo interactions. At the 
present time we do not have sufficient information to disentangle the 
two effects.

\begin{figure*}
\resizebox{0.5\hsize}{!}{\includegraphics{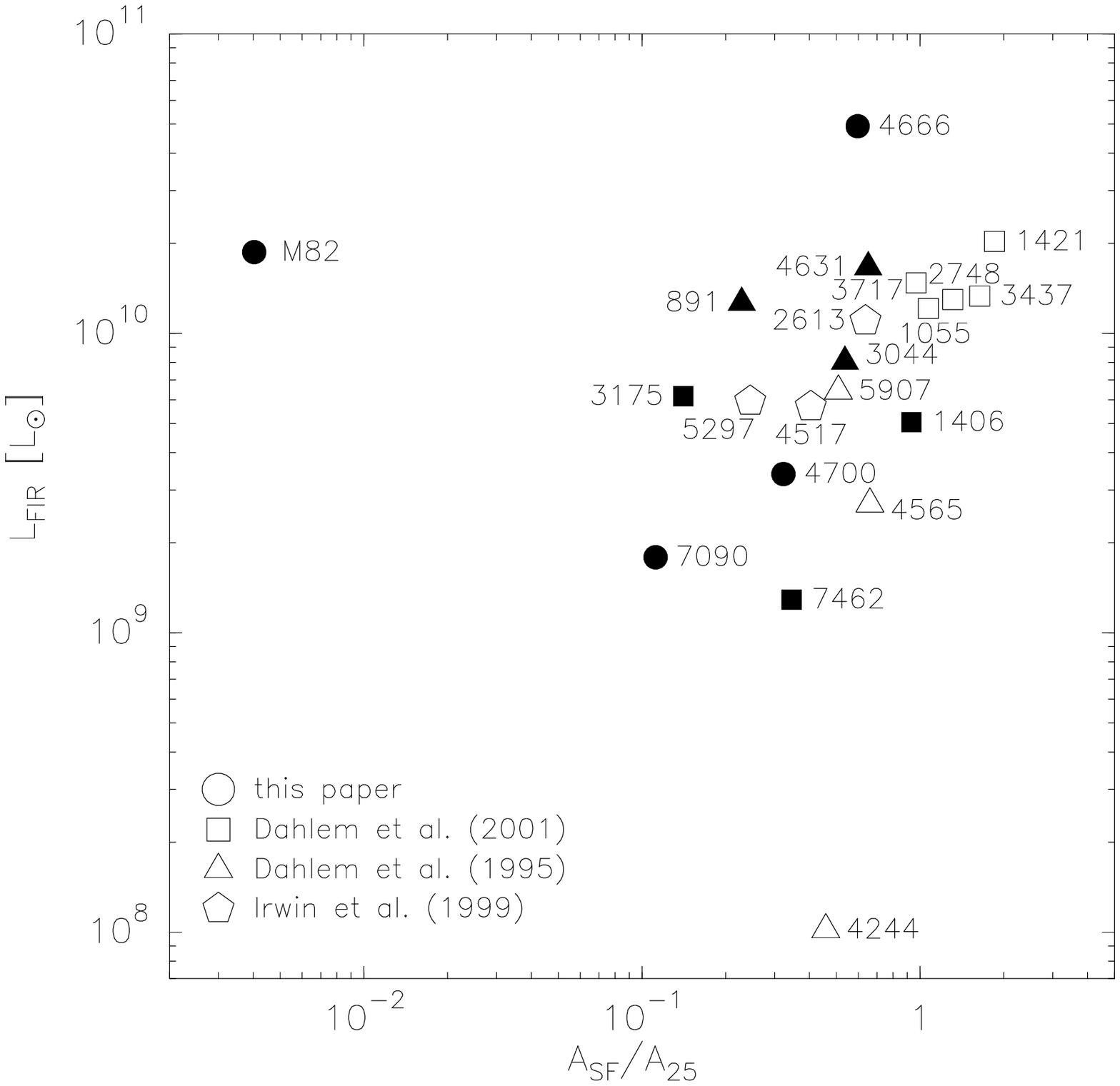}}
\resizebox{0.5\hsize}{!}{\includegraphics{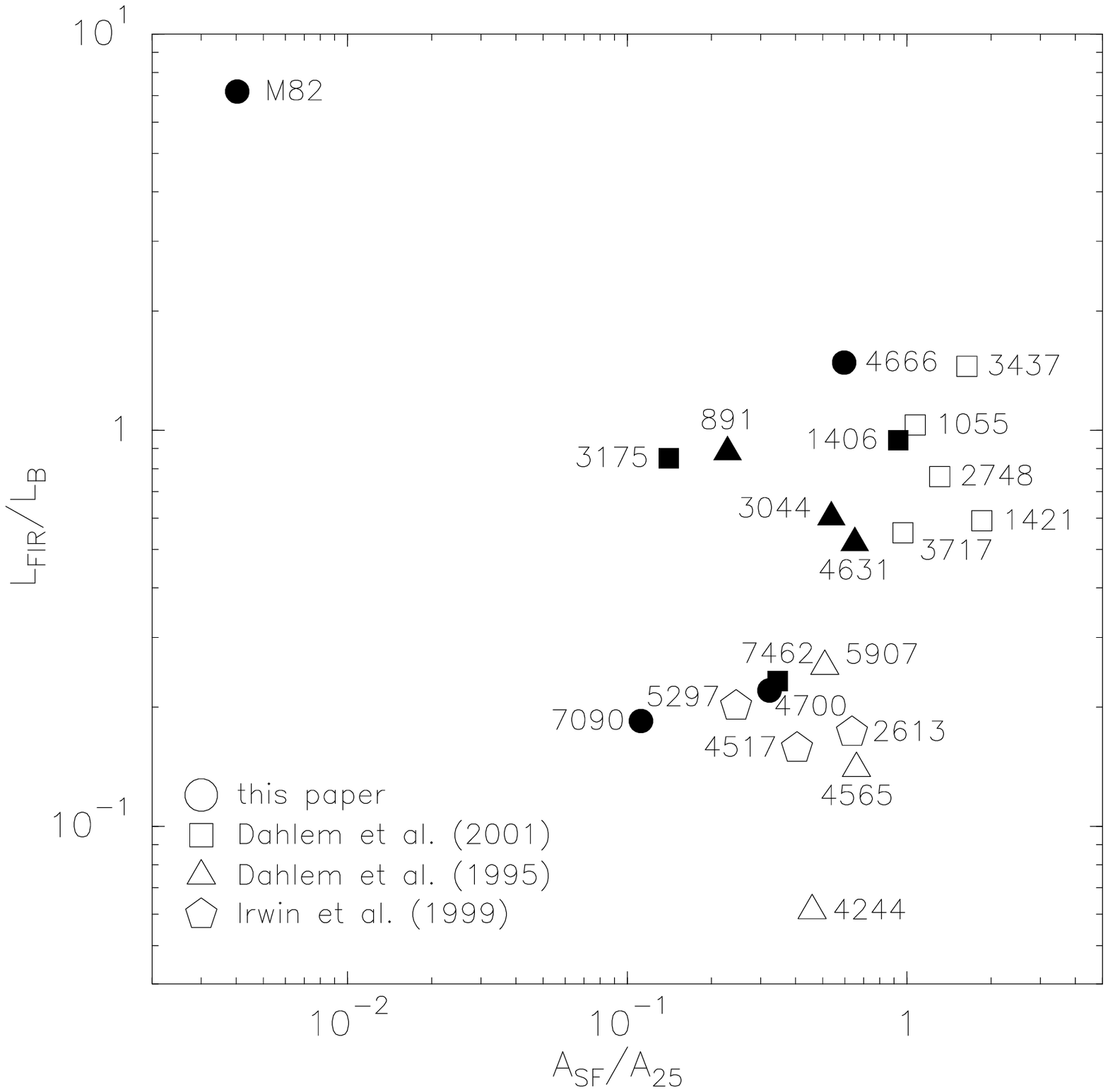}}
\caption{
{\bf Left:} $A_{\rm SF}/A_{25}$ plotted against the far-infrared 
luminosity, \lfir.
{\bf Right:} $A_{\rm SF}/A_{25}$ now plotted against the far-infrared 
vs. optical blue luminosity ratio, \lfir/\lb.
NGC\,1808 is not shown, because the contributions of its AGN and 
starburst to the FIR flux densities could not be separated.
}
\label{fig:radextvslum}
\end{figure*}

\subsubsection{Localised outflows vs. large-scale halos}

The values of \edota\ derived above are averages. These are a measure 
of the global energy input rates into a galaxy's ISM required to form 
large-scale halos.

As calculated by us earlier (DLG95), local values of \edota\ in active 
SF regions are more likely to be of order $10^{-2}$ \ergscm. 
Fig.~\ref{fig:edotavsmass}\ suggests that {\it local} outflows above 
giant \hii\ regions, with \edota\ $\simeq\ 10^{-2}$ \ergscm, should 
always be possible, even in otherwise quiescent galaxies with
high mass surface densities.
This explains why a galaxy such as NGC\,4565, with a low SF rate and 
thereby low average energy input level (see Fig.~\ref{fig:edotavsmass}), 
can exhibit a small number of dust filaments perpendicular to its disk 
plane which might be tracers of localised outflows of material. 
Note that, apart from the non-detection of a radio halo, NGC\,4565 is 
also deficient in H$\alpha$ and [O{\small{\sc III}}] in the disk-halo 
interface (Rand et al. 1992 and Robitaille et al. 2006), which again 
argues against the existence of a large-scale gaseous halo.

The decisive difference between a circumnuclear starburst and an 
individual localised outflow in a disk with more widespread SF is 
the total amount of energy injected into the ISM.
The energy injection rates, \edota, are yet higher in starbursts 
than in Giant Extragalactic \hii\ regions (GEHRs; Kennicutt 1984).
Also, since classical starbursts comprise a number of star-forming 
regions each of which equals a GEHR, energy (and mass) is deposited
into the ambient medium over a longer time than in an individual 
GEHR.
Hence, while for an individual GEHR the total energy can amount to
of order $10^{54}$ ergs, that of a classical starburst is orders of 
magnitude larger, nearer $10^{56}$ ergs. In the case of ultra-luminous 
infrared galaxies (ULIRGs) up to $10^{58-59}$ ergs are deposited 
(HAM90).

\subsubsection{Compactness of the star formation distribution}

The left panel of Fig.~\ref{fig:radextvslum}\ suggests that there 
is no clear link between the far-infrared luminosities, \lfir, and 
the ratio of the surface areas of the star-forming parts vs. the
total extent to 25th optical blue magnitude of the galaxy disks, 
$A_{\rm SF}/A_{25}$, in our sample galaxies. 
In fact, when comparing the position of M\,82 on one hand with 
those of NGC\,891 and NGC\,4631 on the other, one can see that for 
widely different SF distributions galaxies can have the same \lfir.
\footnote{
NGC\,1808 was excluded from this part of the investigation, because 
the contribution of its low-luminosity Seyfert 2 AGN to the total 
FIR luminosity is unknown. In the determination of its radio flux 
density (above) the flux contribution of the AGN was subtracted
by fitting the central peak using the AIPS task {\sc JMFIT}.}

However, in the right panel of Fig.~\ref{fig:radextvslum}, although
again there is no clear correlation between the two quantities, a 
trend is indicated that galaxies with radio halos have smaller 
$A_{\rm SF}$/$A_{25}$ values and at the same time higher 
\lfir/\lb\ ratios than the ones without halos. This suggests that
a compact SF distribution favours the creation of a gaseous halo.

\begin{figure*}
\resizebox{0.5\hsize}{!}{\includegraphics{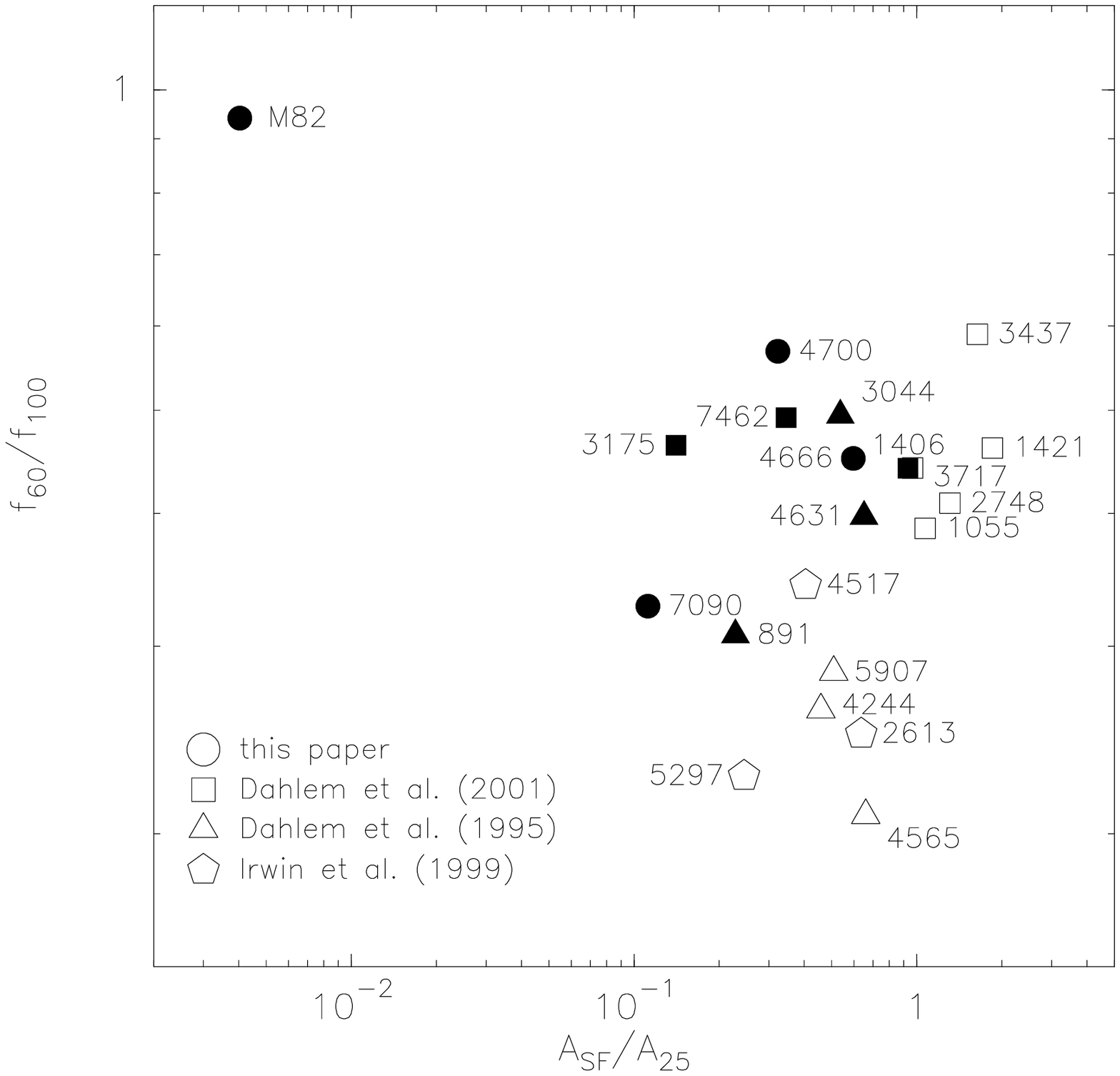}}
\resizebox{0.5\hsize}{!}{\includegraphics{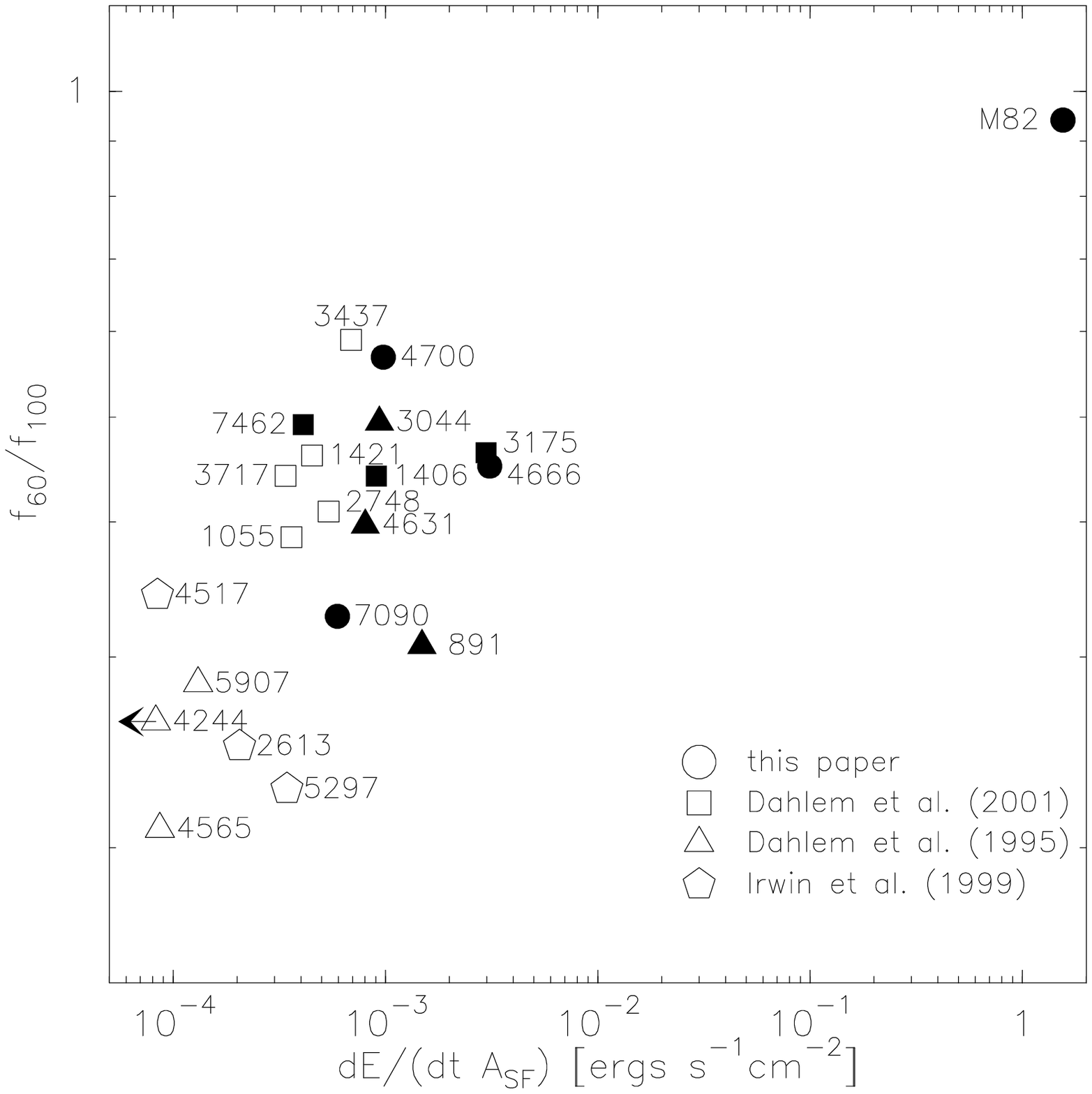}}
\caption{
{\bf Left:} $A_{\rm SF}/A_{25}$ plotted vs. the IRAS 60 $\mu$m to 
100 $\mu$m flux ratio, which is a measure of the average warm dust 
temperature. 
{\bf Right:} Energy input rates per unit surface area, \edota, 
plotted vs. $f_{60}/f_{100}$. 
NGC\,1808 is not shown, because the contributions of its AGN and 
starburst to the FIR flux densities could not be separated.
}
\label{fig:sfvsfircol}
\end{figure*}

The left panel of Fig.~\ref{fig:sfvsfircol}\ shows another trend, 
namely for galaxies with radio halos to have both compact SF 
distributions and warm average dust temperatures (as reflected by 
their high IRAS 60 $\mu$m to 100 $\mu$m flux ratios, $f_{60}/f_{100}$).
The right panel of Fig.~\ref{fig:sfvsfircol}, which is similar to
Figs.~38 and 39 by Dahlem et al. (2001) and Fig.~5 by Rossa and 
Dettmar (2003a), indicates that galaxies with radio halos also
have high energy input rates into their ISM and correspondingly 
high average dust temperatures. 

These relationships show that any combination of high \lfir/\lb\ 
ratio, compact SF distribution and high $f_{60}/f_{100}$ flux ratio
(\gapeq0.4) is a good tracer of galaxies with radio halos.
This explains why high FIR flux densities and high $f_{60}/f_{100}$ 
flux ratios (indicating the presence of warm dust) are such good 
selection criteria to find, after rejection of AGNs, galaxies with 
SF-driven gaseous halos (Dahlem et al. 2001).

\subsubsection{Energy input rate vs. halo z-height}

A search for a dependence of the radio halo $z$-scale heights on 
the level of energy input in the disk did not yield conclusive
results. Since the uncertainties involved are yet too large to 
make firm statements, this topic will not pursued any further at 
this time.

\subsection{Future work}

It is unlikely that large numbers of new targets fulfilling
the stringent selection criteria for studies like this will
be found in the near future. The main search criteria (FIR
flux and colour) become more unreliable when going to more
distant and thus fainter sources. In addition, the detection
and proper imaging of gaseous halos in distant galaxies is
extremely time-consuming on the present generation of 
telescopes. A much larger sample will become available
only through the advent of the next generation of more
powerful telescopes (both optical and radio). Progress in
this field at the present time can be made by obtaining
more complete databases for the galaxies presented here
(and a few others known already, including e.g. NGC\,253)
and by finding ways to measure halo properties in face-on
(or at least low inclination) galaxies.

\acknowledgements{
It is a pleasure to thank the following colleagues for 
contributing data to this study:
S. Ryder for making available an H$\alpha$ image of NGC\,3175;
M. Lehnert for providing us with an H$\alpha$ image of M\,82;
U. Klein for making available the radio continuum images of
M\,82 by Reuter et al. (1992).
UL acknowledges support by the Spanish Ministry of Education,
via the research projects AYA\,2005-0716-C02-01, 
ESP\,2004-06870-C02-02, and the Junta de Andaluc\' ia.
This research has made use of the NASA Extragalactic Database
and the LEDA database, the contributions of which are gratefully 
acknowledged.
We thank the anonymous referee for very constructive criticism
and useful suggestions which have led to significant improvements
of the paper.
The Digitized Sky Survey was produced at the Space Telescope 
Science Institute under U.S. Government grant NAG W-2166. The 
National Geographic Society -- Palomar Observatory Sky Atlas 
(POSS-I) was made by the California Institute of Technology 
with grants from the National Geographic Society. 
}

\end{document}